\documentclass[showpacs,10pt,twocolumn,prl]{revtex4-1}
\usepackage{amsmath}
\usepackage{amssymb}
\usepackage{graphics}
\usepackage{epsfig}
\usepackage{CJK}
\usepackage{color}
\usepackage{multirow}

\begin{document}

\title{Extremely large magnetoresistance and high-density Dirac-like fermions in ZrB$_{2}$}
\author{Qi Wang,$^{\dag}$ Peng-Jie Guo,$^{\dag}$ Shanshan Sun, Chenghe Li, Kai Liu,$^{*}$ Zhong-Yi Lu, and Hechang Lei}
\email{kliu@ruc.edu.cn}
\email{hlei@ruc.edu.cn}
\affiliation{Department of Physics and Beijing Key Laboratory of Opto-electronic Functional Materials $\&$ Micro-nano Devices, Renmin University of China, Beijing 100872, China}

\date{\today}

\begin{abstract}
We report the detailed study on transport properties of ZrB$_{2}$ single crystal, a predicted topological nodal-line semimetal. ZrB$_{2}$ exhibits extremely large magnetoresistance as well as field-induced resistivity upturn and plateau. These behaviors can be well understood by the two-band model with the perfect electron - hole compensation and high carrier mobilities. More importantly, the electrons with small effective masses and nontrivial Berry phase have significantly high density when compared to those in known topological semimetals. It strongly suggests that ZrB$_{2}$ hosts Dirac-like nodal-line fermions.

\end{abstract}


\maketitle



Recently discovered topological semimetals (TSMs), as a new kind of gapless-type topological quantum materials, have induced extensive research interest because of exotic physical phenomena and potential applications for novel devices. TSMs are characterized by the robust bulk band crossings (nodal points) near Fermi energy level $E_{F}$ \cite{Burkov,XuG,WanX,WangZ}. Due to the novel bulk and surface topological band structures, TSMs exhibit many of exotic transport and spectroscopic properties. For example, there are linear dispersions around band crossing points with unusual quasiparticle excitations behaving like Dirac or Weyl fermions in high energy physics \cite{WangZ,WangZ2,LiuZK1,LiuZK2,WengH,XuSY,LvBQ}. Moreover, TSMs host novel topological surface states (SSs), such as the Fermi arc and the drumhead SSs \cite{WanX,XuG,WengH,XuSY,LvBQ,XuSY2,WengH2,YuR,BianG}. They also exhibit extremely large magnetoresistance (XMR) with ultrahigh carrier mobility \cite{LiangT,Shekhar}, and chiral anomalies with negative longitudinal MR \cite{HuangXC,XiongJ}.

According to the momentum space distribution and degeneracy of the nodal points, the TSMs can be classified into two types. The first class of TSMs has zero-dimensional (0D) discrete nodal points, such as fourfold-degenerate Dirac points \cite{WangZ,WangZ2,LiuZK1,LiuZK2} and twofold-degenerate Weyl points \cite{WengH,XuSY,LvBQ} etc. In contrast, there is a line along which two bands cross each other in the Brillouin zone in the second class namely topological nodal-line semimetals (TNLSMs) \cite{Burkov,WengH2}. The one-dimensional (1D) nodal line can connect each other to form nodal ring, nodal chain or nodal net \cite{FangC,Bzdusek}. Many materials have been proposed as the candidates of TNLSMs in theory, such as all-carbon Mackay-Terrones crystals \cite{WengH2}, Cu$_{3}$PdN \cite{YuR}, PbTaSe$_{2}$ \cite{BianG}, IrF$_{4} $\cite{Bzdusek}, Ca$_{3}$P$_{2}$ \cite{XieLS,ChanYH}, ZrSiCh (Ch = S, Se, and Te) \cite{Schoop,HuJ}, and CaAgX (X = P and As) \cite{Yamakage,Takane,WangXB} etc. However, the bulk nodal-ring states and drumhead SSs have been only observed in limited materials, such as PbTaSe$_{2}$, ZrSiCh, and CaAgX etc \cite{BianG,Schoop,Takane,WangXB}. The nodal-chain and nodal-net states are still elusive.

Very recently, first-principles calculations predict that AlB$_{2}$-type diborides MB$_{2}$ (M = Sc, Ti, V, Zr, Hf, Nb, and Ta) could be a new family of TNLSMs, which can host nodal ring and nodal net with threefold-degenerate points when spin-orbital coupling (SOC) is ignored \cite{ZhangX,FengX}.
Moreover, these materials have larger energy ranges of linear dispersion ($>$ 2 eV) than other TSMs \cite{ZhangX,FengX}, in favor of the experimental study. Stimulating by these theoretical studies, in this work, we perform the detailed study on magnetotransport properties of ZrB$_{2}$ single crystals. ZrB$_{2}$ exhibits XMR as well as significant field-induced resistivity upturn and plateau at low temperature. Experimental and theoretical results further indicate that there are two sets of three-dimensional (3D) Fermi surfaces (FSs) with nearly compensated carrier densities. These carriers exhibit high mobilities, light effective masses and nontrivial Berry phase. More importantly, the carrier densities in ZrB$_{2}$ are much higher than the ones in most of TSMs even TNLSMs ZrSiCh.


Single crystals of ZrB$_{2}$ were grown by the Fe flux method (details are presented in the Supplemental Material \cite{SM}). X-ray diffraction (XRD) of a single crystal was performed using a Bruker D8 X-ray machine with Cu $K_{\alpha}$ radiation. Electrical transport measurements were carried out by using Quantum Design PPMS-14T. The longitudinal and Hall electrical resistivity were measured by using a standard four-probe method. The electronic structures of ZrB$_{2}$ were studied by using the first-principles calculations \cite{SM}).


\begin{figure}[tbp]
\centerline{\includegraphics[scale=0.42]{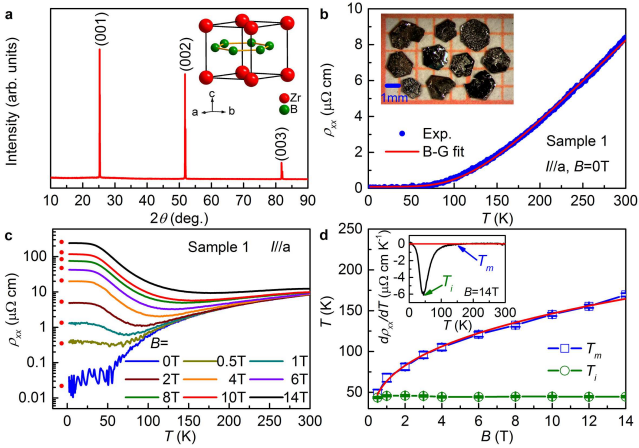}} \vspace*{-0.3cm}
\caption{
(a) XRD of a ZrB$_{2}$ single crystal. Inset: crystal structure of ZrB$_{2}$; (b) Temperature dependence of zero-field $\rho_{xx}(T,0)$ for $I\Vert a$, fitted using the B-G formula. Inset: photo of typical ZrB$_{2}$ single crystals. (c) Temperature dependence of $\rho_{xx}(T, B)$ at various fields. The red solid points are obtained from the formula $\rho_{0}+\alpha B^{2}/\rho_{0}$, where $\alpha$ is fitted from the MR curve at 2 K. (d) Field dependence of $T_{m}$ and $T_{i}$, corresponding to the sign change and the minimum in the $d\rho_{xx}(T, B)/dT$ curves, respectively. The red solid line is the fit using $T_{m}(B)\propto(B-B_{c})^{1/n}$ where $B_{c}=\rho_{0}/\alpha^{1/2}$. Inset: $d\rho_{xx}(T, B)/dT$ vs. $T$ at 14 T. The positions of $T_{m}$ and $T_{i}$ are marked by arrows.}
\end{figure}


ZrB$_{2}$ adopts an AlB$_{2}$-type centrosymmetric structure with the space group $P6/mmm$ (No. 191). The B atoms form graphene-like hexagon planes along the $ab$ plane and Zr atoms are located between two B layers (inset of Fig. 1(a)). The powder XRD confirms the pure phase of ZrB$_{2}$ sample and the fitted lattice parameters are $a=$ 3.1692(1) \AA\ and $c=$ 3.5307(2) \AA\ (Fig. S1 in the Supplemental Material \cite{SM}), close to the reported values in the literature \cite{Branscomb}. The XRD pattern of a ZrB$_{2}$ single crystal indicates that the surface of crystal is the $(00l)$ plane (Fig. 1(a)). The hexagon shape of ZrB$_{2}$ crystal (inset of Fig. 1(b)) is consistent with the single crystal XRD pattern and its crystallographic symmetry. The zero-field in-plane resistivity $\rho_{xx}(T,0)$ of ZrB$_{2}$ single crystal 
exhibits good metallic behavior and the rather large residual resistance ratio [RRR $\equiv \rho_{xx}$(300 K)/$\rho_{xx}$(2 K) $\simeq$ 386] indicates the high quality of crystals (Fig. 1(b)). According to the Bloch-Gr\"{u}neisen (B-G) formula \cite{Ziman},

\begin{equation}
\rho_{xx}(T)=\rho_{0}+C(\frac{T}{\Theta_{D}})^{5}\int^{\Theta_{D}/T}_{0}\frac{x^{5}}{(e^{x}-1)(1-e^{-x})}dx
\end{equation}

\noindent where $\rho_{0}$ is the residual resistivity, $\Theta_{D}$ is the Debye temperature, $C$ is a constant. The good fit of $\rho_{xx}(T)$ curve over the full temperature range (red solid line in Fig. 1(b)) shows that the $e$ - $ph$ scattering is dominant in ZrB$_{2}$. The fitted $\Theta_{D}$ is 779(3) K, close to the values derived from ZrB$_{2}$ polycrystal \cite{Fisher}. Such high Debye temperature can be ascribed to the vibration of B atoms with light mass.

When applying a magnetic field, the $\rho_{xx}(T, B)$ curve exhibits a clearly upturn behavior (Fig. 1(c)). Even at a very small field of 0.5 T, the slope of $\rho_{xx}(T, B)$ curve shows a sign change from positive at high temperature to negative at low temperature, i.e, there is a minimum in the $\rho_{xx}(T, B)$ curve at "turn-on" temperature $T_{m}(B)$ (inset of Fig. 1(d)). With increasing the field, the $T_{m}(B)$ shifts to higher temperature gradually (Fig. 1(d)). On the other hand, there is a plateau in $\rho_{xx}(T, B)$ curve following the upturn behavior at low temperatures and high fields. The temperature at which the resistivity plateau begins to appear seems unchanged with fields. It can be seen more clearly from the field dependence of the characteristic temperature $T_{i}(B)$ related to the inflection point of $\rho_{xx}(T, B)$ curve (inset of Fig. 1(d)). As shown in Fig. 1(d), the $T_{i}(B)$ is about 45 K when $B>$ 1 T and almost insensitive to the field.

ZrB$_{2}$ exhibits significantly large MR [$=(\rho_{xx}(T,B)-\rho_{xx}(T,0))/\rho_{xx}(T,0)\times 100\%$] at low temperature. The MR at 2 K reaches 4.2$\times$10$^{5}$ \% and 1.0$\times$10$^{6}$ \% at 9 T and 14 T for $H\Vert c$ (Fig. 2(a)), comparable to the MRs in other TSMs and compensated semimetals (SMs) \cite{LiangT,HuangXC,Shekhar,Ali,Tafti,SunSS}. Moreover, the MR does not saturate up to 14 T and the Shubnikov-de Haas (SdH) quantum oscillations (QOs) appear at low-temperature and high-field region (Fig. 2(a)). The MR at 2 K can be well fitted using the formula MR $\propto B^{m}$ with $m=$ 1.989(4) (Fig. 2(a)), very close to the typically quadratic field dependence of the MR in the multiband compensated metals \cite{Ziman}. The MR decreases gradually with increasing temperature, but it is still unsaturated up to 14 T even at 300 K.
As shown in Fig. 2(b), when the field rotates away from the $c$ axis towards the $ab$ plane ($\theta=$ 0$^{\circ}$ corresponding to $B\Vert c$), the MR 
at 14 T increases at first and then decreases. Finally, it reaches a minimum value at $\theta=$ 90$^{\circ}$ ($B\Vert ab\perp I$). This complex angular-resolved magnetoresistance (AMR) behavior can be seen more clearly on the polar plot of AMR at 2 K and 14 T (Fig. 2(c)). There are multiple local maximum and minimum values on the plot and the maximum value appears at $\theta\sim$ 26$^{\circ}$. Similar behaviors have been observed previously when the field is relatively low \cite{Piper}. This reflects the large anisotropy of FSs and/or relaxation times. Even the MR is highly anisotropic, the field dependence of unsaturated MR for all of field directions is still close to quadratic (Table S1 and Fig. S2 in the Supplemental Material \cite{SM}). This could be attributed to the (nearly) perfect compensation of electrons and holes. Moreover, the SdH QOs can be observed for all of field directions. When field is tilting from $B\Vert c\perp I$ ($\phi=$ 0$^{\circ}$) to $B\Vert a\Vert I$ ($\phi=$ 90$^{\circ}$), the MR at 2 K and 14 T has a maximum value for $\phi=$ 0$^{\circ}$ and decreases monotonically (Fig. 2(d)). However, there is no negative longitudinal MR observed when $B\Vert I$ ($\phi=$ 90$^{\circ}$). Similar to the MR curves in Fig. 2(b), the derived $m$ of the curves in Fig. 2(d) are also close to 2 (Table S2 and Fig. S3 in the Supplemental Material \cite{SM}). On the other hand, the MR with $I\Vert [210]$ (rotating 90$^{\circ}$ from the $a$ axis in the $ab$ plane) exhibits similar field dependence to those with $I\Vert a$ (Table S3 and Fig. S4 in the Supplemental Material \cite{SM}). Thus, the MR is insensitive to the current direction rotated in the $ab$ plane. In addition, the negative longitudinal MR is still absent when $I\Vert [210]$.

\begin{figure}[tbp]
\centerline{\includegraphics[scale=0.16]{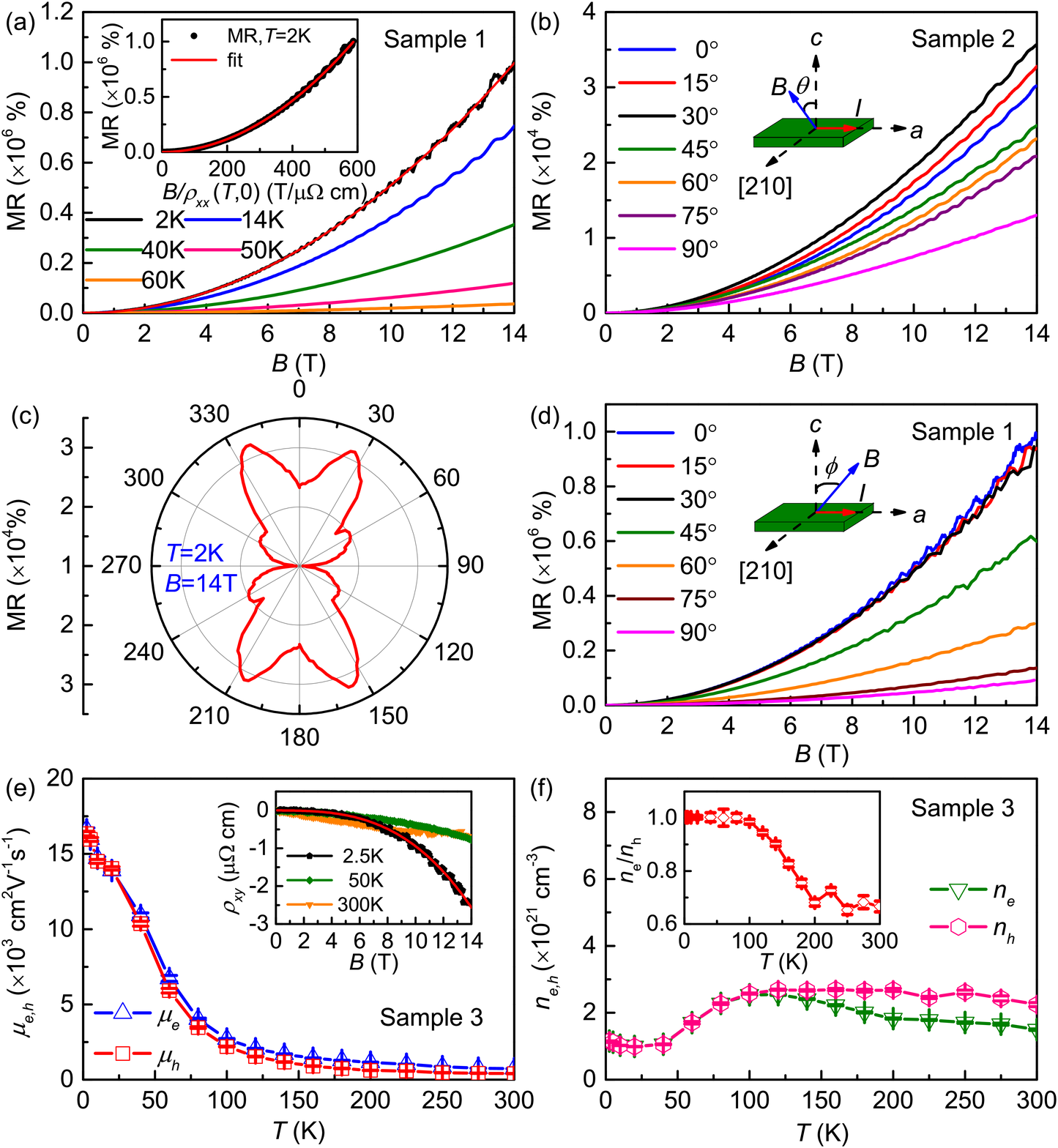}} \vspace*{-0.3cm}
\caption{
(a) Field dependence of MR at various temperatures. The red solid line is the fit using the MR $\propto B^{m}$ at 2 K. Inset: the relation between MR and $B/\rho_{xx}(T,0)$ at $T=$ 2 K. The red solid line shows the fit using MR $=\alpha (B/\rho_{xx}(T,0))^{2}$. (b) Field dependence of MR at 2 K with the field rotating from the $c$ axis to the $ab$ plane. The field is always perpendicular to the current direction ($I\Vert a$). (c) Polar plot of AMR at 2 K and 14 T. (d) Field dependence of MR at 2 K with the field titling from the $c$ axis to the current direction ($a$ axis). (e) Temperature dependence of fitted carrier mobilities $\mu_{e,h}(T)$. Inset: field dependence of $\rho_{xy}(T,B)$ at several typical temperatures. The red solid line is the fit using the two-band model. (f) Temperature dependence of fitted carrier concentrations $n_{e,h}(T)$. Inset: the ratio of $n_{e}/n_{h}$ a function of temperature.}
\end{figure}

Inset of Fig. 2(e) shows the field dependence of Hall resistivity $\rho_{xy}(T,B)$ at various temperatures. The $\rho_{xy}(T,B)$ at high temperature has a nearly linear dependence on field; when decreasing temperature, the $\rho_{xy}(T,B)$ bends downward at high field. The nonlinear behavior clearly indicates that ZrB$_{2}$ is a multiband metal. Moreover, the QOs can be seen at low-temperature and high-field region, consistent with the MR results. Using the two-band model \cite{Ziman},

\begin{equation}
\rho_{xx} = \frac{1}{e}\frac{(n_{h}\mu_{h}+n_{e}\mu_{e})(1+\mu_{e}\mu_{h}B^{2})}{(n_{h}\mu_{h}+n_{e}\mu_{e})^{2}+(n_{h}-n_{e})^{2}(\mu _{e}\mu _{h})^{2}B^{2}}
\end{equation}

\begin{equation}
\rho_{xy} = \frac{B}{e}\frac{(n_{h}\mu_{h}^{2}-n_{e}\mu_{e}^{2})+(n_{h}-n_{e})(\mu_{e}\mu_{h})^{2}B^{2}}{(n_{h}\mu_{h}+n_{e}\mu_{e})^{2}+(n_{h}-n_{e})^{2}(\mu _{e}\mu _{h})^{2}B^{2}}
\end{equation}

\noindent where $\mu_{e,h}$ and $n_{e,h}$ are the mobilities and concentrations of electron- and hole-type carriers, respectively. The $\rho_{xx}(T,B)$ and $\rho_{xy}(T,B)$ can be fitted very well (red solid line in the inset of Fig. 2(e) and Fig. S5 in the Supplemental Material \cite{SM}) and the obtained $\mu_{e,h}(T)$ and $n_{e,h}(T)$ as a function of temperature are shown in Fig. 2(e) and (f). The $\mu_{e,h}(T)$ exhibits similar temperature dependence, i.e., monotonically decreases with similar slopes when increasing temperature (Fig. 2(e)), reflecting similar $e$ - $ph$ scattering mechanism for both types of carriers. At high temperature, $\mu_{e}$ is about twice larger than $\mu_{h}$ and both of them become rather high at low temperature (1.66(1) and 1.64(1)$\times$10$^{4}$ cm$^{2}$ V$^{-1}$ s$^{-1}$ for $\mu_{e}$ and $\mu_{h}$ at 2 K, respectively). On the other hand, when $T<$ 125 K, the $n_{e,h}(T)$ are almost the same and increase gradually with temperature (Fig. 2(f)). The estimated $n_{e}$ and $n_{h}$ at 2 K are 1.156(5) and 1.153(5)$\times$10$^{21}$ cm$^{-3}$. Correspondingly, the ratio of $n_{e}$/$n_{h}$ at 2 K is very close to one (1.003(9)) with tiny amounts of excess $n_{e}$ compared to $n_{h}$ (inset of Fig. 2(f)). This undoubtedly indicates that the carriers in ZrB$_{2}$ are nearly compensated at low temperature. In contrast, when $T>$ 125 K, the $n_{e}(T)$ starts to decrease while the $n_{h}(T)$ still keep increasing until about 200 K, leading to the gradual decrease of the ratio of $n_{e}$/$n_{h}$.

\begin{figure}[tbp]
\centerline{\includegraphics[scale=0.16]{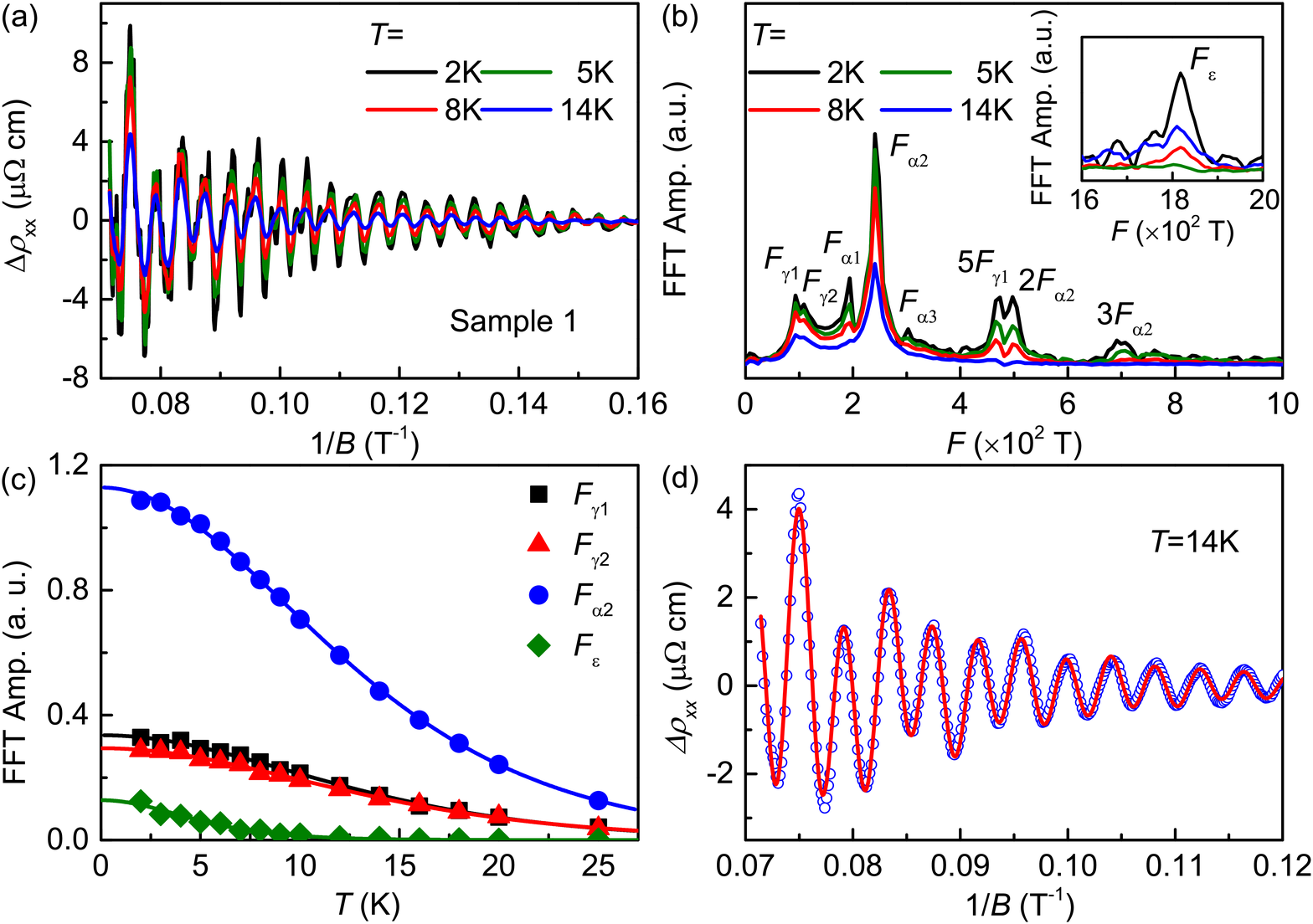}} \vspace*{-0.3cm}
\caption{
(a) SdH QOs $\Delta \rho_{xx} = \rho_{xx}-<\rho_{xx}>$ as a function of $1/B$ at various temperatures for $B\Vert c$. (b) FFT spectra of the QOs at various temperatures. Inset: FFT spectra at high-frequency region. (c) The temperature dependence of FFT amplitudes of $F_{\gamma1}$, $F_{\gamma2}$, $F_{\alpha2}$, and $F_{\varepsilon}$ peaks. The solid lines represent the L-K formula fits for $m^{*}$s. (d) Two-band L-K formula fit (red solid line) of SdH QO (blue empty circles) at $T=$ 14 K.}
\end{figure}

Analysis of SdH QOs provides further insight on the features of FSs and carriers. The oscillation parts of resistivity $\Delta \rho_{xx} = \rho_{xx} - \langle \rho_{xx}\rangle$ against the reciprocal of magnetic field $1/B$ for $B\Vert c$ at few representative temperatures are shown in Fig. 3(a). The amplitudes of QOs exhibit complex periodic behaviors, indicating the contributions of multiple frequency components. They decrease with increasing temperature or decreasing field, but still persist up to about 20 K. The fast Fourier transform (FFT) spectra of SdH QOs for $B\Vert c$ reveal several fundamental frequencies $F_{\gamma1}=$ 93.2 T, $F_{\gamma2}=$ 108.8 T, $F_{\alpha1}=$ 194.2 T $F_{\alpha2}=$ 240.8 T, $F_{\alpha3}=$ 303.0 T, and $F_{\varepsilon}=$ 1817.7 T, and their corresponding higher harmonic frequencies (main panel and inset of Fig. 3(b)). These frequencies are well consistent with the results derived from the de Haas-van Alphen QOs \cite{Tanaka,Pluzhnikov}. According to the Onsager relation $F=(\hbar/2\pi e)A_{F}$, where $A_{F}$ is the area of extremal orbit of FS. The determined $A_{F}$ is 0.0089 - 0.0104, 0.0185 - 0.0289 and 0.1733 \AA$^{-2}$ for the $\gamma$, $\alpha$, and $\varepsilon$ external obits, respectively. The $A_{F}$s of the $\gamma$, $\alpha$ frequencies are relatively small when compared to that of the $\varepsilon$ frequency. The former takes about 0.26 \% - 0.30 \% and 0.54 \% - 0.85 \%, and the latter occupies about 5.09 \% of the whole area of Brillouin zone in the $k_{x}-k_{y}$ plane providing the lattice parameter $a=$ 3.1692 \AA.
In general, the SdH QOs with several frequencies can be described by linear superposition of multi-frequency Lifshitz-Kosevich (L-K) formula, each of which can be expressed as \cite{Shoenberg,HuJ,Lifshitz,Pavlosiuk},

\begin{equation}
\Delta\rho_{xx}^{i}\propto \frac{5}{2}\sqrt{\frac{B}{2F}}R_{T}R_{D}R_{S}\cos[2\pi(F/B+\gamma-\delta+\varphi)]
\end{equation}

\noindent where for the $i$-th SdH QO component, $F$ is frequency, $R_{T}=(\lambda m^{*}T/B)/\sinh(\lambda m^{*}T/B)$, $R_{D}={\rm exp}(-\lambda m^{*}T_{D}/B)$, $R_{S}={\rm cos}(\pi m^{*}g^{*})$, $m^{*}$ effective cyclotron mass in unit of free electron mass $m_{0}$, $T_{D}$ the Dingle temperature, $g^{*}$ effective g-factor, and constant $\lambda=2\pi^{2}k_{B}m_{0}/e\hbar\approx$ 14.7 T/K. The phase factor $\gamma-\delta+\varphi$ contains $\gamma=1/2-\phi_{B}/2\pi$ where $\phi_{B}$ is Berry phase, $\delta$ is determined by the dimensionality of FS ($\delta=$ 0 and $\pm$ 1/8 for the two-dimensional and 3D cases) \cite{Shoenberg,Mikitik,Lukyanchuk}, and $\varphi=$ 1/2 ($\rho_{xx}\gg\rho_{xy}$) or 0 ($\rho_{xx}\ll\rho_{xy}$) \cite{XiangFX}. First, the $m^{*}$s are obtained by fitting the temperature dependence of the FFT amplitudes to the $R_{T}$ \cite{Rhodes}. The fitted $m^{*}$ is 0.112(1), 0.107(1), 0.1126(7) and 0.29(1) $m_{0}$ for $F_{\gamma1}$, $F_{\gamma2}$, $F_{\alpha2}$ and $F_{\varepsilon}$, respectively (Fig. 3(c)).
Setting the obtained $m^{*}$ as known parameter, the precise values of $\phi_{B}$ and $T_{D}$s can be obtained from the fit of SdH QO using multi-frequency L-K formula.
Here we focus on the low-frequency QO components because the relatively low measuring field limits to get reliable values of $\phi_{B}$ and $T_{D}$ for the high-frequency $F_{\varepsilon}$. In addition, in order to minimize the influences of Zeeman splitting and harmonic frequencies, the SdH QO at 14 K is fitted. The two-frequency L-K formula can describe the SdH QO quiet well (Fig. 3(d)). The fitted $T_{D}$ are 47(2) and 20.9(4) K for $F_{\gamma}$ (average of $F_{\gamma1}$ and $F_{\gamma2}$) and $F_{\alpha2}$, corresponding to the quantum mobilities $\mu_{Q}=\frac{e\hbar}{2\pi k_{B}m^{*}T_{D}}=$ 4.2(2)$\times$10$^{2}$ and 9.1(2)$\times$10$^{2}$ cm$^{2}$ V$^{-1}$ s$^{-1}$. The $\mu_{Q}$ is smaller than the $\mu_{e}$ because the former is sensitive to both large- and small-angle scattering, whereas the latter is affected by only large-angle scattering \cite{Shoenberg}. Because of $\rho_{xx}\gg\rho_{xy}$ and a strong 3D character of FSs in ZrB$_{2}$ (shown below), we take $\varphi=$ 1/2 and $\delta=\pm$ 1/8. Based on these values, the fitted $\phi_{B}$ are 0.583(5)$\pi$ ($\delta=$ 1/8) and 1.083(5)$\pi$ ($\delta=$ -1/8) for $F_{\gamma}$, and 0.005(2)$\pi$ ($\delta=$ 1/8) and 0.505(2)$\pi$ ($\delta=$ -1/8) for $F_{\alpha2}$. Thus, there is a nontrivial $\phi_{B}$ for $F_{\gamma}$. The variation of $\phi_{B}$ at different extremal orbits of same electron pocket (shown below) could be related to the anisotropy of FS. Similar behavior has been observed in TNLSM ZrSiCh \cite{HuJ}.

\begin{figure}[tbp]
\centerline{\includegraphics[scale=0.42]{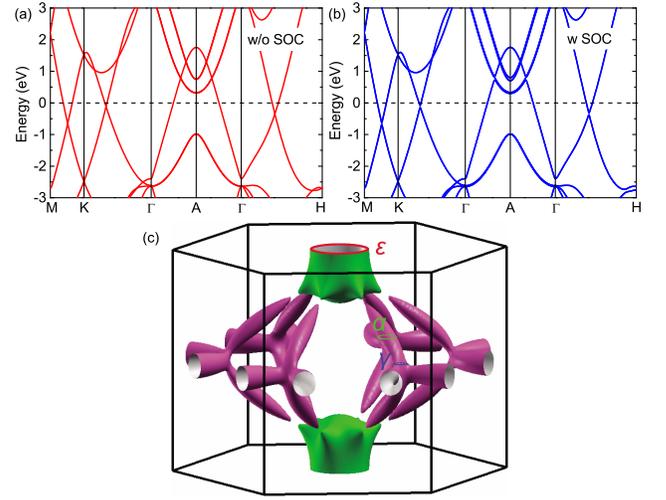}} \vspace*{-0.3cm}
\caption{
Band structure of ZrB$_{2}$ calculated (a) without and (b) with the SOC effect. The horizontal dashed line denotes the position of $E_{F}$. (c) Calculated electron-type FS around $K$ point and hole-type FS around $A$ point. The extremal orbits observed in the SdH QOs are labeled.}
\end{figure}

Theoretical calculations show that there are two bands crossing the $E_{F}$ for ZrB$_{2}$ (Fig. 4(a)). The electron-type FS surrounds the nodal rings in $\Gamma-M-K$ and $\Gamma-K-H$ planes, which connect to each other around $K$ points and form a nodal net (Fig. 4(c)). This is consistent with previous calculations \cite{ZhangX,FengX}. The complex 3D electron FS could be the origin of the strong anisotropy of MR (Fig. 2(c)). When the SOC is included, small energy gaps ($\sim$ 40 - 60 meV) open around the nodal net, but the dispersions of electron band are still linear in a wide energy range up to 3 eV around the $E_{F}$ (Fig. 4(b)), indicating Dirac-like fermions with high carrier mobilities existing in ZrB$_{2}$. The hole pocket is located at $A$ point showing a 3D cylinder-like FS with the long principle axis along $\Gamma-A$ direction. The $\alpha$ and $\gamma$ frequencies in the SdH QO measurements (Fig. 3(b)) can be ascribed to the extremal orbits of electron pockets while the $\varepsilon$ frequency corresponds to that of hole pocket (Fig. 4(c)) \cite{Tanaka,Pluzhnikov}. Based on the volume information of electron and hole pockets, the calculated concentrations of electron- and hole-type carriers $n_{e,h}$ are 1.32 and 1.27$\times$10$^{21}$ cm$^{-3}$, which are in good agreement with the results of Hall measurements. The calculations once again confirm that ZrB$_{2}$ is a compensated semimetal ($n_{e}/n_{h}=$ 1.04) with relatively high carrier densities.


The field-induced resistivity upturn and plateau behaviors have been frequently observed in TSMs and compensated SMs, such as NbP, WTe$_{2}$, LaSb/LaBi and ZrSiS etc \cite{Shekhar,Ali,Tafti,SunSS,Singha}. Several mechanisms have been proposed to explain these behaviors, such as perfect electron-hole compensation \cite{SunSS,WangYL,GuoPJ}, and field-induced gap opening at the Weyl (Dirac) points \cite{Shekhar,Khveshchenko}. The following analysis strongly implies that the large MR and field-induced behaviors in ZrB$_{2}$ can be explained well by the two-band model with electron-hole compensated condition and high carrier mobilities. According to the two-band model with the condition of carrier compensation ($n_{e}=n_{h}$) \cite{Ziman,SunSS,WangYL}, first, the MR equals $=\mu_{e}\mu_{h}B^{2}=\alpha[B/\rho_{xx}$($T$, 0)]$^{2}$. Thus, the field dependence of MR is quadratic (inset of Fig. 2(a)) and the large MR is directly related to the high $\mu_{e,h}$ in ZrB$_{2}$ at low temperature. At high temperature, combined with the quick decreases of $\mu_{e,h}$ and the uncompensated carriers, the MR at high temperature becomes much smaller than that at low temperature.
Second, there should be a minimum in $\rho_{xx}(T,B)$ curve when $B>B_{c}(=\rho_{0}/\alpha^{1/2})$ \cite{SunSS,WangYL}, i.e., field-induced resistivity upturn. Using $\rho_{xx}$($T=$ 2 K, 0) = 0.022 $\mu\Omega$ cm and fitted $\alpha=$ 0.029 ($\mu\Omega$ cm/T)$^{2}$ (inset of Fig. 2(a)), the determined $B_{c}$ is about 0.13 T, well consistent with the resistivity minimum appearing when $B>$ 0.5 T (Fig. 1(c)). Third, a resistivity plateau with the value of $\rho_{0}+\alpha B^{2}/\rho_{0}$ should appear at low temperature \cite{SunSS}. As shown in Fig. 1(c), the derived values of resistivity plateau (red solid points) is in good agreement with the experimental values at 2 K. Fourth, if the $\rho_{xx}(T,0)$ can be described  approximately by the formula $\rho_{0}+AT^{n}$, the $T_{m}(B)$ should be proportional to $(B-B_{c})^{1/n} $ \cite{SunSS}. Providing the $B_{c}$ is 0.13 T, the fitted $n$ is 2.92(8) (Fig. 1(d)), very close to the fitted $n$ (= 2.92(2)) from the $\rho_{xx}(T,0)$ curve between 2 K and 170 K (corresponding to the $T_{m}(B)$ at $B$ = 14 T) (Fig. S6 in the Supplemental Material \cite{SM}).

The high $\mu_{e,h}$ in ZrB$_{2}$ could be partially ascribed to the small $m^{*}$s. Such small $m^{*}$s are comparable with those in the known TSMs \cite{LiangT,Shekhar,HuJ}.
In contrast, the $n_{e,h}$ in ZrB$_{2}$ are much higher than those in the TSMs with discrete nodal points ($\sim$ 10$^{17}$ - 10$^{18}$ cm$^{-3}$) \cite{LiangT,XiongJ}, and even larger than those of TNLSM ZrSiCh ($\sim$ 10$^{20}$ cm$^{-3}$) \cite{HuJ}. Combined with the nontrivial $\phi_{B}$ of electron pocket, it strongly suggests that the electron-type carriers in ZrB$_{2}$ have the feature of Dirac-like nodal-line fermions.




In summary, ZrB$_{2}$ exhibits XMR and field-induced exotic phenomena at low temperature. The nearly perfect electron-hole compensation and remarkably high mobilities $\mu_{e,h}$ are the essential conditions leading to these exotic phenomena. More importantly, ZrB$_{2}$ has very large $n_{e,h}$ with significantly small $m^{*}$s and nontrivial $\phi_{B}$, especially for electron band. This strongly implies that there are Dirac-like nodal-line fermions in ZrB$_{2}$. Even there are high-density carriers, the small $m^{*}$s of Dirac-like fermions can still result in very high $\mu_{e,h}$, a prerequisite for the XMR behavior. Thus, current work will shed light on exploring novel XMR materials in a broader scope, other than semimetals with low carrier concentrations.



This work was supported by the National Key R\&D Program of China (2016YFA0300504, 2017YFA0302903), the National Natural Science Foundation of China (Grant No. 11474356, 11574394, 11774423, 11774424, 91421304), and the Fundamental Research Funds for the Central Universities, and the Research Funds of Renmin University of China (RUC) (14XNLQ03, 15XNLF06, 15XNLQ07, 16XNLQ01). Computational resources were provided by the Physical Laboratory of High Performance Computing at Renmin University of China.

$\dag$ These authors contributed equally to this work.

\end{document}